\documentclass{emulateapj}



\def\***#1{{\sc #1}} \def\plan#1{\relax} \def\Plan#1{\relax}
\def\PLAN#1{\relax}

\def\lta{\mathrel{\spose{\lower 3pt\hbox{$\mathchar"218$}} \raise
2.0pt\hbox{$\mathchar"13C$}}} \def\gta{\mathrel{\spose{\lower
3pt\hbox{$\mathchar"218$}} \raise 2.0pt\hbox{$\mathchar"13E$}}}
\newcommand{\etal}{{\it et al.}}  

\shortauthors{Senorita Devi et al.}
\shorttitle{Extremely luminous X-ray point sources}

\def\mathnew{\mathsurround=0pt}

\def\simov#1#2{\lower .5pt\vbox{\baselineskip0pt \lineskip-.5pt
\ialign{$\mathnew#1\hfil##\hfil$\crcr#2\crcr\sim\crcr}}}

\begin{document}

\title{The dependence of the estimated luminosities of ULX on spectral models }

\author{A. Senorita Devi\altaffilmark{1},  R. Misra\altaffilmark{2},  V. K. Agrawal\altaffilmark{3} and K. Y. Singh\altaffilmark{1} }

\altaffiltext{1}{Department Of Physics, Manipur University, Canchipur,
Imphal-795003, Manipur, India; senorita@iucaa.ernet.in}

\altaffiltext{2}{Inter-University Center for Astronomy and
Astrophysics,  Post Bag 4, Ganeshkhind, Pune-411007, India;
rmisra@iucaa.ernet.in}

\altaffiltext{3}{Department of Astronomy and Astrophysics,
Tata Institute of Fundamental Research, Homi Bhabha Road, Mumbai-400 005,India}

\begin{abstract}
Data from {\it Chandra} observations of thirty nearby
galaxies were analyzed and 365 X-ray point sources were
chosen whose  spectra were not contaminated
by excessive diffuse emission and not affected
by photon pile up. The spectra of these sources
were fitted using two spectral models
(an absorbed  power-law and a disk blackbody) to ascertain the
dependence of estimated parameters on the spectral model used. It was found 
that the cumulative luminosity function depends on the choice of the
spectral model, especially for luminosities $> 10^{40}$ ergs/s.
In accordance with previous results, a large number ($\sim 80$) of the
sources have luminosities $> 10^{39}$ ergs/s (Ultra-Luminous X-ray
sources) with indistinguishable average
spectral parameters (inner disk temperature $\sim 1$ keV and/or photon
index $\Gamma \sim 2$)  with those of the lower luminosities ones. 
After considering foreground stars and known background AGN,
we identify four sources  whose minimum luminosity
exceed $10^{40}$ ergs/s, and call them  Extremely Luminous X-ray
sources (ELX). The spectra of these sources 
are in general better represented by the disk black body model 
than the power-law one. These ELX can be grouped into two distinct
spectral classes. Two of them have an inner disk temperature
of $< 0.5$ keV and  hence are called ``supersoft'' ELX, while the other two
have temperatures $\gtrsim 1.3$ keV and are called
``hard'' ELX.  The estimated inner disk temperatures of the
supersoft ELX are compatible with the hypothesis that they
harbor intermediate size black holes, which are accreting
at $\sim 0.5$ times their Eddington Luminosity. The radiative
mechanism for hard ELX, seems to be Inverse Comptonization, which
in contrast to standard black holes systems, is probably saturated.  
Extensive variability analysis of these ELX, will be able to distinguish
whether these two spectral class represent different systems or
they are spectral states of the same kind of source. 

\end{abstract}


\section{Introduction} 

In the last few years, {\it Chandra} observations of nearby galaxies
have detected many non-nuclear X-ray point sources 
\citep{Kaa01,Mat01,Zez02}, 
some of which
have isotropic luminosities $> 10^{39}$ ergs/s and are
called Ultra luminous X-ray Sources (ULX). 
While some of these ULX are 
 supernova remnants \citep[e.g.][]{Ryd93,Fox00}, it
 is believed that the majority of them are compact accreting systems.
 Indeed, ASCA X-ray spectral studies of many ULX have revealed that they
 display the characteristics of accreting black holes 
\citep{Mak00,Miz01}.
 ULX have also been called Super-Eddington
 Sources \citep{Fab89,Fab04} and Intermediate luminosity X-ray
 objects \citep{Rob00,Col99}.

 Since
these ULX sources emit radiation at a rate larger than the Eddington
luminosity for a ten-solar mass black hole, they are believed to
harbor a black hole of mass $10 \, M_\odot\! < \! M \! <\! 10^5
\,M_\odot$ \citep{Col99,Mak00}
where the upper limit 
is constrained by the fact that a
more massive black hole would have settled into the nucleus due to
dynamical friction \citep{Kaa01}. 
Black holes in this mass range are called 
Intermediate Mass Black Holes (IMBH), since they
 seem to represent the missing component of the black hole
mass spectrum with masses prevailing in the gap between those of
stellar mass black holes found in Galactic X-ray binaries and those associated
with  Active Galactic Nuclei, $ M \sim 10^{6}-10^{9}$ $M_{\odot}$
\citep{Ric98}. \cite{Mil04} and \cite{Mil05} review the present evidence 
for IMBH in ULX and \cite{Liu05} have compiled a catalogue of some ULX
and properties.

Alternate models for ULX are that their luminosities are
super-Eddington \citep{Beg02} or that their emission is
beamed from a geometrically thick accretion disk \citep{Kin01}.
However, it has been argued that in the latter case, such thick
"funnel" shaped disks enhance the observed flux by just a factor of
few \citep{Mis03}. For all of these models, the creation of such
sources \citep{Por02,Tan00,Mad01} and process by which they sustain
high accretion rates \citep{Kin01}, are largely unknown.

Investigations on the nature of ULX have been undertaken by studying
the spectra and variability of individual sources. For example 
analysis of the spectra of  NGC 1313 X-1, X-2 \citep{Mil03} and
M81 X-9 \citep{Mil04a}, revealed 
the presence of a cool accretion disk component ($kT_{in} \sim 0.1-0.5 $ keV),
suggesting that ULX indeed harbor IMBH. 
Transitions between two spectral states, similar to those seen in Galactic
black hole systems, have been reported in  NGC 1313 X-1\citep{Col99} 
and two sources in IC342 \citep{Kub01}. Spectral transitions have also
been reported in two sources in NGC 1313 \citep{Fen06}.

The large collecting area of XMM-Newton,
allows for detailed spectral fits to ULX, which often comprise of two
components \citep{Wan04,Fen05}.   However, \cite{Gon06}, have
argued that such soft spectral components depend on the complexity
of the fitting model. An interesting object is
the brightest X-ray point source in M 82, whose intrinsic luminosity
has been measured to be as high as  1.6 $\times$ 10$^{41}$ ergs/s 
\citep{Pta99}. The detection of a $54$~mHz quasi-periodic oscillation 
in its X-ray light curve 
suggests that the source is a compact object 
and not a background AGN \citep{Str03,Dew06a}. The spectra of this source can
be fitted by a power-law with photon index, $\Gamma \sim 2$ \citep{Fio04}, 
but is more consistently fitted with a flatter power-law with
an high energy cutoff around $\sim 6$ keV, which can be interpreted as
optically thick, saturated Comptonization \citep{Agr06}. A 
quasi-periodic oscillation has also been discovered in the bright X-ray
source of Holmberg IX, which is similar to the source in  M82 in having a
flat spectrum ($\Gamma \sim 1$)  with a $\sim 9$ keV cutoff \citep{Dew06}. 
Recently, \cite{Sto06} found that the XMM-Newton spectra of  eleven of the 
eighteen ULX studied by them,  showed such high energy curvature. {\it Chandra}
observations of NGC 5204 X-1, also reveals the presence of an optically
thick Comptonized component \citep{Rob06}.
While these results of individual ULX are intriguing, there does not seem
to be any significantly distinguishable spectral property of ULX,
and in general
their spectra can be described either by steep or flat power-law indices,
with and without soft components \citep[e.g.][]{Dew05}.

Another line of investigation is to construct the cumulative luminosity 
function and histograms of spectral parameters of a large sample of
X-ray sources \citep[e.g.][]{Col02}. 
The hope here is that, in case ULX are a distinct class
of sources and/or they can be classified into distinct subgroups,
the luminosity function should exhibit a break and their spectral parameters
should show clustering. \cite{Swa04} analyzed data from 82 galaxies and
estimated the luminosity function and spectral parameters of the X-ray
point sources in these galaxies. They found that the average photon
index (as well as  the distribution) of the ULX  and the less luminous
sources is nearly same. Moreover, their spatial
and variability distributions are also similar. While, their analysis
revealed that the luminosity functions of ULX depend on the host
galaxy type and star formation rate, they did not find any
significant evidence for breaks in them.

As emphasized by \cite{Swa04}, the power-law model they used to fit the
data was chosen as an empirical one.  They attempted
to fit all the sources with the power-law model and only for those sources
that did not provide a reasonable fit, they used other models like disk
black body. They note that for many sources that are well 
fitted by the power-law model, other spectral models can  
also represent the spectra.
In this work, we consider a smaller sample of 30 galaxies
but fit the spectra of the points sources, with both a power-law
and a disk blackbody model. In principle, the intrinsic 
(i.e. the absorption corrected) luminosity  inferred
for a source may be different for the two spectral models. Our 
first motivation here is to make a qualitative estimate of this difference by noting
the dependency of the luminosity function on the spectral model used.
A second motivation for this work is based on the expectation that estimations of a 
different spectral parameter like the inner
disk temperature (as compared to the photon index), maybe better in
distinguishing ULX from other sources or they may  reveal dependencies
(like correlations between luminosity and temperature) 
which could shed light on the nature of these sources. As mentioned earlier,
the spectra of several ULX are complex requiring more than one components.
Thus, the two spectral models, an absorbed power-law and and an absorbed 
disk blackbody, should be considered as  empirical ones, which can adequately
represent low count data and these models need not be 
the correct physical representation of the actual source spectrum.

It is clear that ULX are fairly common in nearby galaxies. For example,  
\cite{Swa04} identified 154 of them in 82 galaxies. If these sources are
Eddington limited, then black holes $\ga 10 M_\odot$ may indeed be 
quite common.
However, only a few of these sources have luminosities greater than
$10^{40}$ ergs/s. It is these sources, (the best example being
the bright source in M82 X-1), that  require black hole with masses, 
$M  \sim 10^{2-4} M_\odot$. The  development of a self-consistent theory
which explains the process by which such black holes
are created and undergo high accretion, is theoretically challenging.
 Hence, it is important to
 estimate the number of sources
whose minimum intrinsic luminosity exceeds $10^{40}$ ergs/s. Here
the minimum value of the luminosity should not only include the
statistical spectral fitting error, but also the variations in
the luminosity estimation that may occur upon  using 
different viable spectral models. The third motivation of this work is 
to identify such sources 
which we call Extremely Luminous X-ray sources (ELX). To avoid
possible ambiguities in determining the luminosity, we have chosen only 
those sources whose spectra
are not contaminated by excessive diffuse emission and which
are not affected by photon count pile up. Identification of such
relatively ``clean'' systems would allow for more detailed studies
of their properties to be undertaken and would be the 
first step toward understanding  their nature.

A similar analysis has been undertaken recently 
by \cite{Win06} for thirty galaxies observed by XMM-Newton. Since
their motivation was to check if ULX also exhibit soft and hard spectral
states like black hole binaries, they limited their analysis to bright sources
where detailed spectral analysis could be done. They found 16 sources as possible low-state
ULX and 26 as high state ones. For the high state ULX, the observed range for the black body 
temperature was $0.1-1.0$ keV, with the more luminous sources having lower temperature.
The {\it Chandra} analysis undertaken here, can be compared with this contemporary work
and as we discuss later, the analysis are more or less consistent with each other. This 
consistency is important, since although the larger collecting area of XMM-Newton,
allows for more complex spectral analysis, the higher angular resolution
of {\it Chandra} ensures that a source spectrum is not contaminated by diffuse emission
and flux from nearby sources.

\section{Observations and Data Reduction}

The names of the thirty host galaxies and the details of the
 {\it Chandra} ACIS observations are tabulated in Table 1.
 Distances to the galaxies were obtained 
from \cite{Swa04} and references therein. This sample
of galaxies is a subset of those analyzed by \cite{Swa04}.
 No particular criterion was imposed on the selection, since our motivation
is limited to obtaining enough sources and not to study dependency on
galaxy type.

\begin{deluxetable*} {lcccr}
\tablewidth{40pc} 
\tablecaption{Sample Galaxy properties} 
\tablehead{
\colhead{Galaxy} & \colhead{Distance (Mpc)} & \colhead{ObsID} &
\colhead{$T_{exp}(ks)$} & \colhead{N($\ge 60cts$)}} 
\startdata 
NGC0253 & $2.6$ & $969$ & $13.98$ & $13$ \\  
NGC0628 & $9.7$ & $2058$ &$46.16$ & $7$ \\  
NGC0891 & $10.0$ & $794$ & $50.90$ & $14$ \\
NGC1291 & $8.9$ & $795$ & $39.16$ & $14$ \\  
NGC1316 & $17.0$ & $2022$ & $29.85$ & $9$ \\ 
NGC1399 & $18.3$ & $319$ & $55.94$ & $36$ \\
NGC1569 & $2.2$ & $782$ & $96.75$ & $16$ \\ 
NGC2403 & $3.1$ & $2014$ &$35.59$ & $4$ \\ 
NGC3034 & $3.9$ & $361$ & $33.25$ & $5$ \\ 
NGC3079 &$15.6$ & $2038$ & $26.57$ & $5$ \\ 
NGC3379 & $11.1$ & $1587$ & $31.52$& $7$ \\ 
NGC3556 & $14.1$ & $2025$ & $59.36$ & $15$ \\ 
NGC3628 &$10.0$ & $2039$ & $57.96$ & $14$ \\ 
NGC4125 & $24.2$ & $2071$ & $64.23$ & $8$ \\ 
NGC4365 & $20.9$ & $2015$ & $40.42$ & $9$ \\ NGC4374
& $17.4$ & $803$ & $28.47$ & $4$ \\ NGC4449 & $3.7$ & $2031$ & $26.59$
& $12$ \\ NGC4485/90 & $7.8$ & $1579$ & $19.52$ & $9$ \\ NGC4559 &
$10.3$ & $2027$ & $10.70$ & $1$ \\ NGC4579 & $21.0$ & $807$ & $33.90$
& $3$ \\ NGC4594 & $9.6$ & $1586$ & $18.51$ & $18$ \\ NGC4631 & $7.6$
& $797$ & $59.21$ & $12$ \\  NGC4649 & $16.6$ & $785$ & $36.87$ & $23$
\\ NGC4697 & $11.8$ & $784$ & $39.25$ & $19$ \\ NGC5055 & $9.2$ &
$2197$ & $27.99$ & $16$ \\ NGC5128 & $4.0$ & $962$ & $36.50$ & $22$ \\
NGC5194/5 & $8.4$ & $1622$ & $26.80$ & $18$ \\ NGC5457 & $7.0$ &
$2065$ & $9.63$ & $4$ \\ NGC5775 & $26.7$ & $2940$ & $58.21$ & $15$ \\
NGC6946 & $5.5$ & $1043$ & $58.28$ & $16$ \\
\enddata
\tablecomments{($T_{exp}$)the exposure time in ks; (N) the number of
point sources with total counts from the source$\ge$ 60 as detected by
{\it{wavdetect}} with fluxscale$= 1$} 
\end{deluxetable*}

The data reduction and analysis were done using CIAO3.2 and
HEASOFT6.0.2. Using the CIAO source detection tool {\it wavdetect},
X-ray point sources were extracted from the level 2 event list. 
It was found that
at least 60 counts are required to fit the spectral data with a two parameter
model and hence  only those
sources with net counts $ \ge 60$ were chosen for the spectral analysis.
Choosing a lower threshold of $50$ counts resulted in a large number of
sources for which spectral parameters could not be constrained.
To avoid photon pile up effects, a conservative threshold of
the count rate being $> 0.05$ counts/s was imposed which led to the
rejection of fifteen sources which have been listed in Table 3 of the appendix.
 For some sources, typically near the
nucleus, it was difficult to find nearby source free background regions
and hence these sources were also not included in the analysis. Sources
embedded in excessive diffuse emission (i.e. when the background flux 
was larger than $2$ counts/arcsec$^2$ ) were also rejected.  
Typically, this amounted to considering only those sources  
for which the estimated background counts were less than $20$.
For each data set, observation-specific bad pixel
lists were set in the {\it ardlib} parameter file. Using a combination
of CIAO tools and calibration data, the source and background spectra were
extracted. 

These selection criteria makes the sample incomplete both in
low and  high luminosity ends. Thus the results obtained should not
be used for quantitative analysis of the luminosity functions. The motivation
here is evaluate dependency on spectral models and to identify
sources which have high intrinsic luminosity. Thus, care has been
taken to avoid possibly contaminated data, even if such criteria result
in a loss of sources.

Spectral analysis was done using XSPEC version 12.2 and the data
was fitted in the energy range 0.3-8.0 keV. All  sources were fitted
with two spectral models, the absorbed power-law and
an absorbed disk blackbody. Absorption was taken into account
using the XSPEC model {\it phabs}. Since the number of counts in
each spectrum was typically low, the C-statistics was used for the
analysis. Technically, the C-statistics is not appropriate for
high counts and/or for background subtracted data. However, it
was ascertained that the model parameters obtained either by C-statistics
or $\chi^2$ statistics, were consistent with each other for high count
rate sources. This is reassuring, because if the results depended on
the statistics used, it would be imperative to use the correct statistics
for high count sources (which need not be $\chi^2$) 
taking into account the correct (possibly non-Gaussian) error profiles.

An important problem, when fitting low count data with a two parameter
(plus normalization) model, is the possibility of many local minima
in the discerning statistic (in this case C-statistic) space. Hence,
we take a cautious approach and do not fit the data using the 
XSPEC minimization
routine. Instead we compute the C-statistic for a range of parameter
values (using the XSPEC command {\it steppar}) and find the global
minimum. Such a technique is numerically expensive, but it ensures
that the global minimum has been found and the correct errors are
obtained for the best fit parameters.

\section{Results}

The 365 sources considered in this analysis were classified into
three categories, depending on whether the data was better fit by the
disk black body model (23 sources) or the power-law one (67 sources) or 
both (275 sources). The criterion 
chosen to determine a better fit to the data was that C-statistic difference
between the models should be larger than 2.7. If the difference was less,
than both model fits were considered to equally represent the data. Although
such a criterion is ad hoc (considering the uncertainties in the actual
error statistics) and count rate dependent, it does serve as an qualitative 
guideline
to differentiate between those systems which can be represented by
a power-law emission and/or a black body one. The spectral parameters
for all sources for the power-law and disk blackbody models are tabulated
in Tables 4 and 5  of the appendix.

\begin{figure}
\begin{center}
{\includegraphics[width=1.0\linewidth,angle=0]{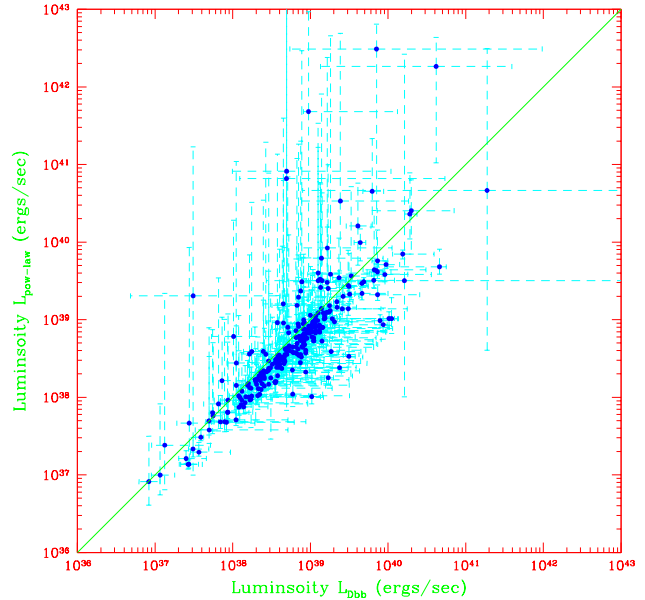}}
\end{center}  
\caption{The intrinsic luminosity (in $0.3$-$8.0$ keV energy range) 
estimated by fitting the power-law model versus the bolometric 
intrinsic luminosity estimated using the disk black body model.}
\end{figure}
For ULX, the physically relevant parameter is the intrinsic bolometric
luminosity which should be used to define and identify them. However,
given the limitations of an instrument's energy sensitivity range, the
bolometric luminosity is spectral model dependent. For a power-law 
the bolometric luminosity cannot be estimated and only a lower limit
can be obtained using the observed energy range. Since our motivation
is to show how the bolometric luminosity is affected by the use of
different spectral models, we have plotted in Figure 1, the bolometric luminosity
for the disk black body model versus the lower limit to the 
luminosity using the power-law model. The figure represents only those sources which can be
represented by both models. The figure shows that while for most sources
the difference in luminosities is not substantial, there are sources 
with estimated luminosities $ \gtrsim 10^{39}$ erg/s, where 
the disk black body luminosity estimation is significantly smaller than
the power-law one.  This happens for sources for which the spectral
index is large.

To show the dependence of the luminosity function on the fitting model,
we compute the cumulative luminosity function in two ways. For the disk 
black body cumulative luminosity function (DBCLF), the 
luminosity obtained form fitting
a black body is used, except for those sources that are fitted better with
a power-law for which the power-law model estimated luminosity is considered. 
Similarly, for the power-law cumulative luminosity function, (PLCLF)
the luminosity corresponds to the power-law fit, except for those sources which
are better represented by a disk black body spectrum. 
In Figure 2, the solid line represents the DBCLF while the
PLCLF is plotted as a dotted one. There are 
less number of sources with $L > 10^{40}$ ergs/s for disk black body preferred
representation. Moreover, there is a significant difference in the slope of
the two luminosity function and the presence of a faint break at
$L \sim 10^{40}$ ergs/s for the PLCLF, is not evident in the DBCLF.
Of particular interest is the number of sources whose minimum luminosity
(i.e. the minimum of the two lower limits obtained by fitting the two
spectral models) exceeds a certain value. In Figure 2,
the dashed line represents such a minimum cumulative luminosity function,
which reveals that there are eight  sources with minimum luminosity 
greater $10^{40}$ ergs/s (ELX) and $\sim 80$ sources with 
 minimum luminosity 
greater $10^{39}$ ergs/s (ULX).
\begin{figure}
\begin{center}
{\includegraphics[width=1.0\linewidth,angle=0]{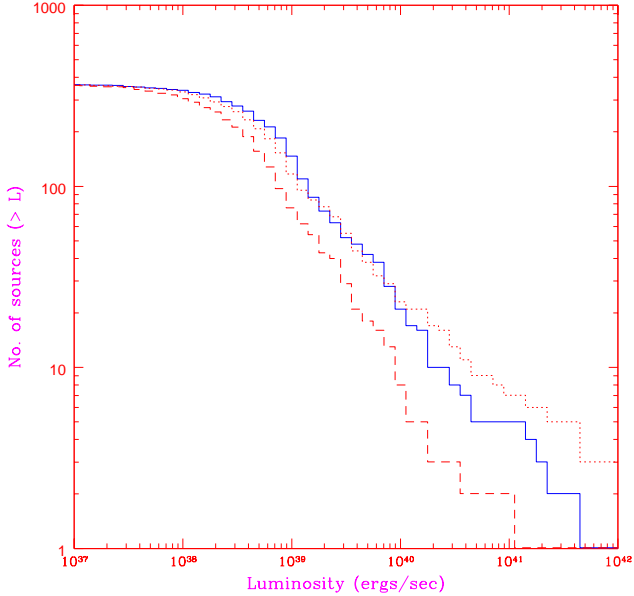}}
\end{center}  
\caption{The cumulative luminosity function using the disk black body
model (solid line), the power-law model (dotted line) and the minimum
cumulative luminosity function (dashed line).}
\end{figure}

\begin{figure}
\begin{center}
{\includegraphics[width=1.0\linewidth,angle=0]{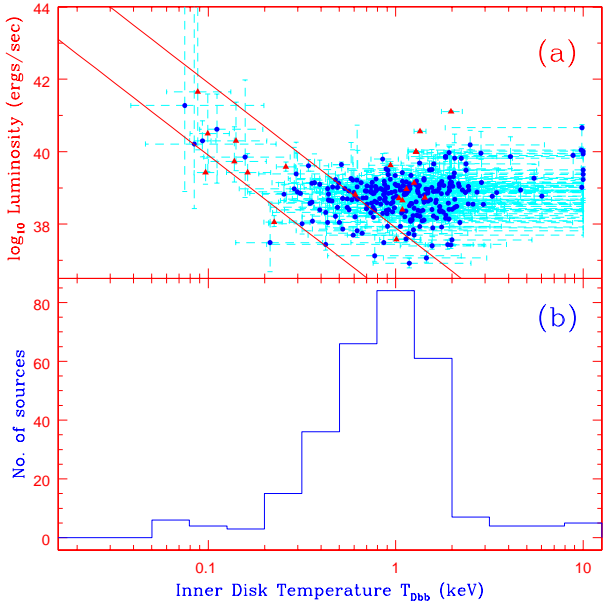}}
\end{center}  
\caption{(a) The luminosity versus inner disk temperature and (b) the
distribution of the inner disk temperature for sources whose spectra can be
modeled as disk black body emission. The triangles represent 
sources which are better fitted by the disk black body model as compared to
the power-law one. The two solid lines represent the expected luminosity
versus maximum temperature relations for accretion disks radiating at one
 and one-tenth of the the Eddington Luminosity. Two sources which were
identified as foreground stars (see text) are not included in this plot.  }
\end{figure}

Figures 3 (a) shows the variation of the luminosity versus 
the disk black body temperature, 
While most of the sources have an inner disk temperature
$\sim 1$ keV, as is evident from the distribution (Figure 3 b), 
there is a population of high luminosity source with 
low ($\sim 0.1$ keV) temperature. Although the number of sources is low,
there seems to be some evidence, that ELX (i.e. sources with luminosities
$> 10 ^{40}$ ergs/s ) can be divided in two groups,  a ``super-soft'' group with temperature
less than  $0.2$ keV and an harder group with temperature $\sim 2$ keV.
Figure 3 may be compared with the results obtained by \cite{Win06} using
XMM-Newton data for a different sample and selection criteria. They also
find that sources with luminosities $> 5 \times 10 ^{39}$ ergs/s have
have a similar  bimodal distribution in temperature as shown in Figure 3.
This supports the hypothesis that ELX can be divided into two groups and
this is not an artifact of sample selection bias.

\begin{figure}
\begin{center}
{\includegraphics[width=1.0\linewidth,angle=0]{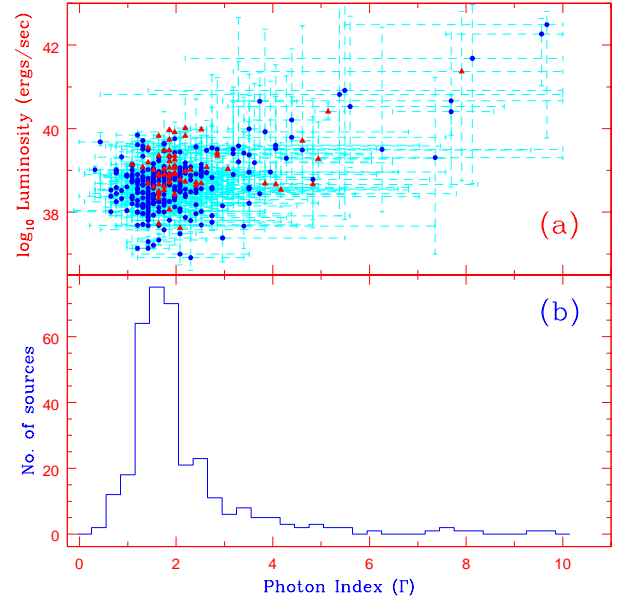}}
\end{center}  
\caption{(a) The luminosity versus power-law index and (b) the
distribution of the power-law index for sources whose spectra can be
modeled as a power-law emission. The triangles represent 
sources which are better fitted by the power-law model as compared to
the disk black body  one. Two sources which were
identified as background AGN (see text) are not included in this plot.}
\end{figure}
Figure 4 (a) shows the variation of luminosity with power-law photon index
for those sources which can be fitted by a power-law model. There is
a clear correlation between the two. This correlation does not seem to be due
to overestimation of column density, since no such correlation is seen
in the luminosity versus $N_H$ plot (Figure 5 b). Similar to the analysis
using the disk black body model, there is a group of ``super-soft'' sources
(i.e.  photon spectral index $> 3$) 
which are also highly luminous ($L > 10^{40}$ 
ergs/s).
The column density versus luminosity plots for both the power-law
and disk black body models (Figure 5), 
reveal an absence of correlation, which is
indicative that there may not be a bias in the analysis, i.e. the luminosities
are not being over-estimated because of a $N_H$ overestimation.
\begin{figure}
\begin{center}
{\includegraphics[width=1.0\linewidth,angle=0]{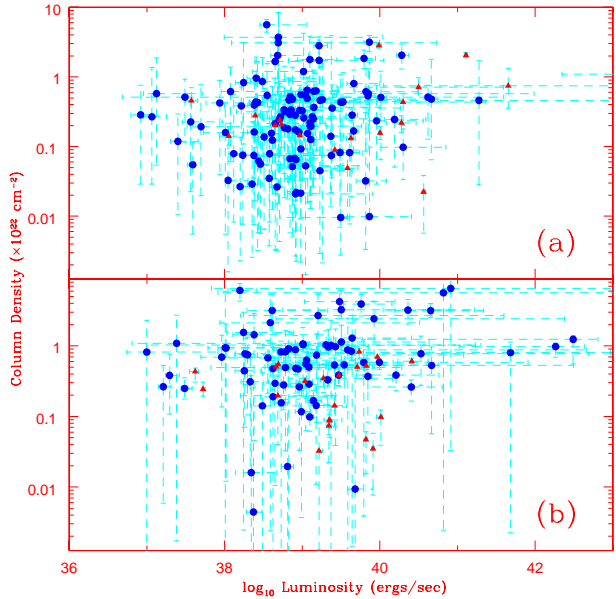}}
\end{center}  
\caption{ The column density versus luminosity for sources fitted with
(a) the disk black body model and (b) the power-law model.}
\end{figure}
In this analysis, there are eight sources  which have an apparent 
minimum luminosity
greater than $10^{40}$ ergs/s. However, two of these sources
(NGC 5055, R.A: 13 15 30.18, Dec: +42 03 13.5 and  NGC 4594, 
 R.A: 12 39 45.22, Dec: -11 38 49.8) are foreground stars based on the 
optical images of the galaxies. Optical spectroscopy of a source in NGC 5775 
( R.A: 14 53 55.8, Dec: +3 33 28.02) reveals that it is a background AGN
\citep{Gut05}, while a source in NGC 1399 ( R.A: 14 53 55.8, Dec: +3 33 28.02)
is a BLAGN \citep{Gre04}. 
The spectral properties of the other four sources, which we call Extremely
Luminous sources (ELX) are tabulated in Table 2. 
The NGC 0628 source reported in Table 2, is a different source than
the well studied ULX, CXOU J013651.1+154547. The luminosity of
this highly variable source \citep{Kra05} is $\sim 10^{39}$ ergs/s and
its spectral properties are listed in Table 4. The source in 
NGC 6946 is a well known variable source \citep{Liu05} and has been
called X7  with $L \sim 10^{39.22}$ ergs/sec\citep{Lir00}, IXO 85 \citep{Col02} and source no. 56 \citep{Hol03}.
Although, the
best fit parameters for C-statistics are shown, these results
have been checked using $\chi^2$ statistics and by C-statistics fit for
unbinned and background not subtracted data and it was found that the
parameters values are consistent within errors and the errors on the
estimated luminosities do not vary by more than a factor two. 
In general, these sources
are better represented by disk black body emission than a power-law model,
except for the source in NGC 6946, which however requires an exceptionally
large power-law photon spectral index, ($\Gamma > 5$).
The spectral properties of these bright sources suggest that they
may be divided into two groups. The first group of four sources
(Table 2), are represented by low inner disk temperatures ($< 0.5$ keV),
and hence may be called ``supersoft'' sources. 
In contrast the second group of three sources, have harder spectra
with inner disk temperatures $\gtrsim 1$ keV or with power-law
photon index ($\Gamma \sim 2$) and hence may be called hard sources.
For this group, the spectra are marginally fitted better with
a disk blackbody emission, although considering the uncertainties in
the spectral fitting, a power-law representation may also be acceptable.

\begin{deluxetable*} {lcccccccccc}
\tablewidth{0pc}
\tablecaption{Spectral properties of sources with minimum luminosity $> 10^{40}$ergs/s }
\tablehead{
\colhead{galaxy} & \colhead{R.A.} & \colhead{Dec.} & \colhead{Counts} & \colhead{Bins} & \colhead{kT$_{in}$ (keV)}
& \colhead{log $L_{Dbb}$}& \colhead{C-stat$_{Dbb}$}& \colhead{$\Gamma$}& \colhead{log $L_{Pow}$}& \colhead{C-stat$_{Pow}$}}
\startdata

NGC0628 &  01 36 47.45 & +15 47 45.01 & 
$200$ & 
$7$ & 
$ 0.09^{+0.04}_{-0.03}$  & 
$41.65^{+2.49}_{-1.60}$ & 
$0.7$ & 
$9.56^{+0.44}_{-2.42}$ & 
$42.18^{+0.36}_{-1.49}$ & 
$8.6$  \\

NGC6946 & 20 35  0.13 & +60  9  7.97 & 
$1936$ &
$66$ &
$ 0.32^{+0.02}_{-0.03}$ &
$39.46^{+0.08}_{-0.06}$ &
$162.2$ &
$ 5.14^{+0.34}_{-0.27}$ & 
$40.41^{+0.21}_{-0.18}$ &
$134.8$ \\

&&&&&&&&&&\\

NGC4579 & 12 37 40.30 & +11 47 27.48 & 
$1696$ &
$66$ &
$ 1.35^{+0.10}_{-0.09}$ &
$40.56^{+0.02}_{-0.02}$ &
$65.0$ &
$ 1.75^{+0.11}_{-0.11}$ &
$40.37^{+0.00}_{-0.02}$ &
$74.5$  \\

NGC5775 & 14 53 58.90 & + 3 32 16.78 & 
$1358$ &
$60$  &
$ 1.97^{+0.29}_{-0.22}$ &
$41.11^{+0.02}_{-0.02}$ &
$55.8$ &
$ 1.86^{+0.22}_{-0.11}$ & 
$40.95^{+0.10}_{-0.06}$ &
$60.1$ \\

\tablecomments{Host Galaxy name; Right Ascension; Declination; Total counts; Number of energy bins after rebinning; Best fit inner disk temperature; 
Bolometric luminosity estimate using disk black body model; C-statistic for 
disk black body model; Photon index $\Gamma$ for the power-law model; 
Luminosity estimate (0.3-8.0 keV) using the power-law model; C-statistic for 
power-law model;}
\enddata
\end{deluxetable*}

\section{Summary and Discussion}

Chandra observations of thirty galaxies were analyzed and the spectra
of their points sources were fitted using both a power-law and a
disk black body emission model. Only those sources were chosen, which were
bright enough to allow a meaningful spectral fit, but whose data
was not contaminated by excessive diffuse emission and/or effected
by pile-up. It was found that the shape of the luminosity function
especially at the high luminosity end, depends on the choice of the
spectral model. 

In accordance with earlier results, a large number of the sources
($\sim 80$ ) have a luminosity which exceeds $10^{39}$ ergs/s and
hence satisfy the standard definition of being Ultra Luminous
X-ray sources (ULX) and do not seem to have any 
spectral distinction when compared with
sources having lower luminosity.  In this sample of 365 sources,
there are four source which we refer to as Extremely luminous X-ray sources
(ELX) since their luminosities were estimated to 
exceed $10^{40}$ ergs/s. These sources are in general better described
by disk blackbody emission and can be distinctively grouped into two 
classes. This is consistent with the results of an independent analysis using
XMM-Newton data \citep{Win06}. The members of the 
first class have soft spectra with best fit inner disk temperature 
$< 0.5$ keV, while for the other class the spectra is harder with 
inner disk temperature $\gtrsim 1.3$ keV.

If disk
black body emission is indeed the correct radiative process for 
the supersoft class then the inner disk temperature should correspond
to  the maximum color temperature of a disk, which can be 
estimated to be
\begin{equation}
T_{col} \sim 0.3 \;\;\hbox {keV} \;\; L_{40}^{-1/4}  (\frac{f}{1.7}) (\frac{L}{L_{Edd}})^{-1/2}
\end{equation}
where $L_{40}$ is the luminosity in $10^{40}$ ergs/s,
$f$ is the color factor and $L_{Edd}$ is the Eddington luminosity. In Figure
3 (a), the two solid lines represent this relationship for 
$L/L_{Edd} = 1$ and $0.1$. Thus within the uncertainties, the supersoft
sources are compatible with having pure black body disk emission, and
have $L \sim 0.5 L_{Edd}$.  

 ELX which are members of the hard class
 have inner disk temperatures which are 
higher than that expected from a Eddington limited black body accretion disks.
Hence, for these source the radiative mechanism is probably inverse
Comptonization of soft photons. Detailed spectral analysis, which
included XMM observations, of the bright X-ray source in M82 X-1 has
revealed that its spectrum is better fitted by a saturated Comptonization
model \citep{Agr06}, which is also the case for the the bright X-ray source
in Holmberg IX. Holmberg IX is not part of the sample studied
here and  
M82 X-1 has been excluded because of pile-up effects and 
excess diffuse emission. Thus, these sources, with estimated luminosities $\sim 10^{41}$ erg/s,
 could also be members
of the hard class of ELX.  Thus it seems that like the 
the hard state of standard black hole binaries, the
hard class ELX also 
have spectra which is due to thermal Comptonization, however 
unlike black hole binaries, in ELX the Comptonization seems to be saturated. 
Thus, it is tempting to draw by analogy, that the two spectral classes of
ELX are actually two spectral states of the same kind of object. This
can be verified if spectral transition between the two classes is observed.

With the identification of these ELX and other sources from the
literature, it is now possible to undertake a more extensive study
of their properties. Temporal variability of these sources will
shed more light on the nature of these enigmatic sources.

\acknowledgements

The authors thank the referee for useful comments and suggestions which have
significantly improved the paper. ASD thanks CSIR and IUCAA for support.

\begin{appendix}

\begin{deluxetable*}{lccr}
\tablewidth{0pc} \tablecaption{List of sources with count rate greater than $0.05$ counts/s} \tablehead{
\colhead{Galaxy} & \colhead{R.A.} & \colhead{Dec.} &
\colhead{Count rate s$^{-1}$} } \startdata 
NGC0253  &  0 47 32.97 & -25 17 48.80 & 0.0845164 \\
NGC0253  &  0 47 22.59 & -25 20 50.87 & 0.0657110 \\
NGC0253  &  0 47 17.55 & -25 18 11.18 & 0.0569807 \\
NGC1569  &  4 31 16.85  & +64 49 50.13   & 0.0567468   \\
NGC2403  &  7 36 55.61 & +65 35 40.85  & 0.0746877 \\
NGC2403  &   7 36 25.53 & +65 35 40.02 & 0.1521207 \\
NGC3628 &  11 20 15.75 & +13 35 13.70 &  0.0530390 \\
NGC4374 &  12 25 11.92 & +12 51 53.53 &  0.0563400 \\
NGC4449 &  12 28 17.83 & +44  6 33.86 & 0.0525013 \\
NGC4485 &  12 30 43.26 &  +41 38 18.36 & 0.0509909 \\
NGC4485 &  12 30 30.56 & +41 41 42.33 &  0.0758634 \\
NGC4559 &  12 35 58.56 & +27 57 41.91 & 0.1233310 \\
NGC4559 &  12 35 51.71 & +27 56  4.05 & 0.1984267 \\
NGC4631 &  12 41 55.56 & +32 32 16.90 &  0.0554584 \\
NGC6946 &  20 35  0.74 & +60 11 30.74 & 0.1448118 \\
\enddata
\tablecomments{The spectra of these sources would be affected by pile-up and hence have not
been included in the sample }
 
\end{deluxetable*}

\begin{deluxetable} {lccccccr}
\tabletypesize{\footnotesize}
\tablewidth{0pc}
\tablecaption{
Spectral Properties of point sources fitted with the Power-Law model}
\tablehead{
\colhead{Galaxy} & \colhead{R.A.} & \colhead{Decl.} & \colhead{$n_H$($10^{22}cm^{-2}$)} & \colhead{$\Gamma$} &  \colhead{log(L) ergs/s} & \colhead{C$_{stat}$} & \colhead{d. o. f.}}
\startdata
NGC0253 &   0 47 43.07 & -25 15 29.28 & $ 0.15^{+ 0.53}_{- 0.15}$ & $ 2.52^{+2.31}_{-1.21}$ & $37.70^{+1.20}_{-0.18}$ & $  9.43$ & $   4$ \\
NGC0253 &   0 47 42.80 & -25 15  2.02 & $ 0.77^{+ 0.51}_{- 0.49}$ & $ 1.86^{+0.66}_{-0.77}$ & $38.26^{+0.27}_{-0.12}$ & $  3.36$ & $   4$ \\
NGC0253 &   0 47 35.25 & -25 15 11.53 & $ 0.48^{+ 0.05}_{- 0.13}$ & $ 2.19^{+0.22}_{-0.33}$ & $38.65^{+0.05}_{-0.08}$ & $ 38.94$ & $  19$ \\
NGC0253 &   0 47 34.28 & -25 17  3.32 & $ 5.66^{+ 5.98}_{- 5.66}$ & $ 5.38^{+4.62}_{-4.95}$ & $40.82^{+3.28}_{-2.90}$ & $  3.21$ & $   2$ \\
NGC0253 &   0 47 34.00 & -25 16 36.51 & $ 1.04^{+ 0.11}_{- 0.10}$ & $ 2.19^{+0.22}_{-0.22}$ & $39.01^{+0.08}_{-0.07}$ & $ 10.93$ & $  27$ \\
NGC0253 &   0 47 33.55 & -25 18 16.51 & $ 0.00^{+ 0.18}_{- 0.00}$ & $ 1.64^{+0.99}_{-0.44}$ & $37.33^{+0.15}_{-0.10}$ & $  2.37$ & $   2$ \\
NGC0253 &   0 47 32.05 & -25 17 21.43 & $ 3.17^{+ 2.20}_{- 2.46}$ & $ 1.97^{+1.10}_{-1.10}$ & $38.61^{+0.72}_{-0.32}$ & $  1.63$ & $   3$ \\
NGC0253 &   0 47 30.98 & -25 18 26.23 & $ 1.55^{+ 1.46}_{- 1.55}$ & $ 1.75^{+1.21}_{-1.21}$ & $38.24^{+0.65}_{-0.21}$ & $  1.76$ & $   3$ \\
NGC0253 &   0 47 28.01 & -25 18 20.21 & $ 0.93^{+ 1.13}_{- 0.92}$ & $ 2.08^{+1.76}_{-1.54}$ & $38.02^{+1.03}_{-0.23}$ & $  2.13$ & $   1$ \\
NGC0253 &   0 47 25.20 & -25 19 45.22 & $ 0.00^{+ 0.15}_{- 0.00}$ & $ 1.31^{+0.66}_{-0.22}$ & $37.96^{+0.09}_{-0.09}$ & $  5.10$ & $   4$ \\
NGC0253 &   0 47 18.50 & -25 19 13.94 & $ 0.00^{+ 0.15}_{- 0.00}$ & $ 1.31^{+0.44}_{-0.22}$ & $37.97^{+0.07}_{-0.05}$ & $  3.94$ & $   5$ \\
NGC0253 &   0 47 40.66 & -25 14 11.71 & $ 0.69^{+ 0.72}_{- 0.56}$ & $ 2.41^{+1.98}_{-1.32}$ & $37.96^{+1.11}_{-0.27}$ & $  1.07$ & $   3$ \\
NGC0253 &   0 47 17.65 & -25 18 26.45 & $ 0.11^{+ 0.49}_{- 0.11}$ & $ 1.31^{+1.54}_{-0.77}$ & $38.12^{+0.27}_{-0.12}$ & $  3.13$ & $   3$ \\
NGC0628 &   1 36 51.06 & +15 45 46.86 & $ 0.03^{+ 0.05}_{- 0.03}$ & $ 1.86^{+0.22}_{-0.11}$ & $39.22^{+0.04}_{-0.02}$ & $ 50.58$ & $  36$ \\
NGC0628 &   1 36 47.45 & +15 47 45.01 & $ 0.89^{+ 0.11}_{- 0.33}$ & $ 9.56^{+0.44}_{-2.42}$ & $42.18^{+0.36}_{-1.49}$ & $  8.56$ & $   4$ \\
. . . .  &   . . . . .  &  & & & & &  \\
\enddata
\tablecomments{Host galaxy name; Right Ascension; Declination; $n_H$, equivalent hydrogen column density; $\Gamma$, photon power-law index; L, X-ray luminosity in the energy range: 0.3-8.0 keV; C-statistics; degree of freedom.
The complete version of this table is in the electronic edition of
the Journal.  The printed edition contains only a sample. }

\end{deluxetable}

\begin{deluxetable} {lccccccr}
\tabletypesize{\footnotesize}
\tablewidth{0pc}
\tablecaption{Spectral Properties of point sources fitted with the disk black body model}
\tablehead{
\colhead{Galaxy} & \colhead{R.A.} & \colhead{Decl.} & \colhead{$n_H$($10^{22}cm^{-2}$)} & \colhead{kT$_{in}$ (keV)} & \colhead{log(L) ergs/s} & \colhead{C$_{stat}$} & \colhead{d. o. f.}}
\startdata
NGC0253 &  0 47 43.07 & -25 15 29.28 & $  0.00^{+ 0.32}_{-0.00}$ & $ 0.59^{+0.42}_{-0.25}$ & $37.69^{+0.37}_{-0.00}$ & $  9.41$ & $   4$ \\
NGC0253 &  0 47 42.80 & -25 15  2.02 & $  0.44^{+ 0.36}_{-0.32}$ & $ 1.56^{+1.95}_{-0.52}$ & $38.40^{+0.26}_{-0.08}$ & $  3.95$ & $   4$ \\
NGC0253 &  0 47 35.25 & -25 15 11.53 & $  0.24^{+ 0.07}_{-0.07}$ & $ 1.04^{+0.21}_{-0.15}$ & $38.71^{+0.04}_{-0.03}$ & $ 35.79$ & $  19$ \\
NGC0253 &  0 47 34.28 & -25 17  3.32 & $  3.08^{+ 5.36}_{-3.07}$ & $ 0.62^{+9.38}_{-0.36}$ & $38.69^{+2.04}_{-0.60}$ & $  3.37$ & $   2$ \\
NGC0253 &  0 47 34.00 & -25 16 36.51 & $  0.64^{+ 0.12}_{-0.10}$ & $ 1.34^{+0.19}_{-0.16}$ & $39.06^{+0.03}_{-0.03}$ & $  8.60$ & $  27$ \\
NGC0253 &  0 47 33.55 & -25 18 16.51 & $  0.00^{+ 0.08}_{-0.00}$ & $ 0.86^{+0.69}_{-0.29}$ & $37.49^{+0.20}_{-0.06}$ & $  3.85$ & $   2$ \\
NGC0253 &  0 47 32.05 & -25 17 21.43 & $  2.04^{+ 1.82}_{-1.62}$ & $ 2.07^{+4.71}_{-0.80}$ & $38.68^{+0.40}_{-0.09}$ & $  1.37$ & $   3$ \\
NGC0253 &  0 47 30.98 & -25 18 26.23 & $  0.96^{+ 0.96}_{-0.86}$ & $ 2.00^{+8.00}_{-0.91}$ & $38.41^{+0.73}_{-0.12}$ & $  1.84$ & $   3$ \\
NGC0253 &  0 47 28.01 & -25 18 20.21 & $  0.62^{+ 0.77}_{-0.56}$ & $ 1.20^{+8.80}_{-0.55}$ & $38.08^{+0.90}_{-0.12}$ & $  2.51$ & $   1$ \\
NGC0253 &  0 47 25.20 & -25 19 45.22 & $  0.00^{+ 0.06}_{-0.00}$ & $ 1.20^{+0.86}_{-0.33}$ & $38.12^{+0.21}_{-0.09}$ & $  7.71$ & $   4$ \\
NGC0253 &  0 47 18.50 & -25 19 13.94 & $  0.00^{+ 0.05}_{-0.00}$ & $ 1.59^{+0.73}_{-0.41}$ & $38.22^{+0.15}_{-0.09}$ & $  6.36$ & $   5$ \\
NGC0253 &  0 47 40.66 & -25 14 11.71 & $  0.42^{+ 0.47}_{-0.38}$ & $ 0.90^{+2.76}_{-0.40}$ & $37.94^{+0.38}_{-0.18}$ & $  1.64$ & $   3$ \\
NGC0253 &  0 47 17.65 & -25 18 26.45 & $  0.08^{+ 0.32}_{-0.07}$ & $ 1.24^{+8.76}_{-0.65}$ & $38.23^{+1.04}_{-0.13}$ & $  2.85$ & $   3$ \\
NGC0628 &  1 36 51.06 & +15 45 46.86 & $  0.00^{+ 0.00}_{-0.00}$ & $ 0.89^{+0.08}_{-0.11}$ & $39.42^{+0.02}_{-0.04}$ & $119.44$ & $  36$ \\
NGC0628 &  1 36 47.45 & +15 47 45.01 & $  0.75^{+ 0.58}_{-0.39}$ & $ 0.09^{+0.04}_{-0.03}$ & $41.65^{+2.49}_{-1.60}$ & $  0.66$ & $   4$ \\
. . . .  &   . . . . .  &  & & & & &  \\
\enddata
\tablecomments{Host galaxy name; Right Ascension; Declination; $n_H$, equivalent hydrogen column density; $kT_{in}$, inner disk temperature; L, Bolometric X-ray luminosity; C-statistics; degree of freedom.
The complete version of this table is in the electronic edition of
the Journal.  The printed edition contains only a sample. }

\end{deluxetable}




\end{appendix}
\end{document}